\documentstyle{l-aa}
\begin{document}
\def\kms{km\thinspace s$^{-1}$~}
\def\ergcs{erg cm$^{-2}$ s$^{-1}$}
\def\aa{A\&A}
\def\aas{A\&ASS}
\def\aj{AJ}
\def\ajs{AJSS}

\title{The Observational Status of Stephan's Quintet
\thanks{Based on data obtained at the 1.5m telescope of the Estaci\'{o}n de
Observaci\'{o}n de Calar Alto (EOCA), Instituto Geogr\'{a}fico Nacional,
which is jointly operated by the Instituto Geogr\'{a}fico Nacional
and the Consejo Superior de Investigaciones Cient\'{\i}ficas through
the Instituto de Astrof\'{\i}sica de Andaluc\'{\i}a}}

\author{
M.~Moles \inst{1,2} \thanks{Visiting Astronomer, German-Spanish
Astronomical Center, Calar Alto, operated by the Max-Planck-Institut
fur Astronomie jointly with the Spanish National Commission for Astronomy}
\and
I.~M\'arquez \inst{3,4}
\and
J.W. Sulentic \inst{5}
}

\offprints{M.~Moles, moles@oan.es }

\institute{
        Instituto de Matem\'aticas y F\'{\i}sica Fundamental,
CSIC, C/ Serrano 121, 28006 Madrid, Spain
\and
        Observatorio Astron\'omico
Nacional, Apdo. 1143, 28800 Alcal\'a de Henares, Madrid, Spain
\and
        Instituto de Astrof\'{\i}sica de Andaluc\'{\i}a (C.S.I.C.),
Apdo. 3004, 18080 Granada, Spain
\and
        Institut d'Astrophysique de Paris, 98 bis Bd Arago, 75014 Paris,
France
\and
        Department of Physics and Astronomy, University of Alabama,
Tuscaloosa, USA 35487
}

\date{Received,  ; accepted,}

\maketitle
\markboth{Moles et al.: Stephan's Quintet}{}

\begin{abstract}
We present new photometric and spectroscopic data for the galaxies in the
compact group known as Stephan's Quintet. We find the strongest evidence
for dynamical perturbation  in the spiral component  NGC~7319. Most of
the damage was apparently caused by nearby NGC~7320C which passed through the
group a few $\times$10$^8$ years ago. NGC~7318B is a spiral galaxy that
shows evidence consistent with being in the early stages of a collision with
the group. NGC~7317 and 18A are either elliptical galaxies or the stripped
bulges of former spiral components. They show no evidence of past or present
merger activity but are embedded in a luminous halo which suggests that
they are interacting with the other members of the group. The low redshift
galaxy NGC~7320 is  most likely a late type, dwarf spiral projected along
the same line of sight as the interacting quartet.

\keywords{Galaxies - Dynamics - Interaction - Compact Groups}

\end{abstract}

\section{Introduction}

Stephan's Quintet (SQ: also known as Arp319 and VV228) is  one of the most remarkable
groupings of galaxies  on the sky. It was the first compact galaxy group discovered and is
certainly the most intensely studied object of that class. After every new observational
effort unexpected aspects have emerged, adding to the puzzling nature of such apparently
dense galaxy aggregate. SQ can be regarded as a prototype of the compact group class (it
is H92 in the catalog of Hickson (1982).

The puzzles connected with SQ began when its component galaxy redshifts were
first measured (Burbidge \& Burbidge 1961). First, one of the four higher
redshift galaxies (NGC~7318B) was found to have a velocity almost 1000 \kms~
lower than the other three raising questions about the groups dynamical
stability. Then, NGC~7320 was found to show a redshift $\sim$5700 \kms~
lower than the mean of the other four. SQ became one of the key objects in
the debate about the nature of the redshift (Sulentic 1983).

It was suggested that NGC~7320 might be a physical companion of NGC~7331
a large, nearby Sb galaxy with similar redshift (\cite{vdb61,a73}). Possible
signs of this hypothesized interaction include a tidal tail that extends from
NGC~7320 towards the SE and an  HI deficiency noted by Sulentic and Arp
(1983). At the same time Arp (1973) interpreted the tail as a sign of
interaction between NGC~7320 and the higher redshift members of SQ.  The
question of the distance to the different galaxies in the group was
specifically addressed by \cite{bbgh73} and \cite{sh74} who tried to use
HI data to settle the question. The results turned out to be contradictory
and later observations (\cite{as80}) showed that the HI in SQ was displaced
from the optical galaxies. A summary of the many papers dealing with
distances to the SQ galaxies can be found in Sulentic (1983).

This paper is concerned with the properties of the five galaxies that comprise
SQ. We consider both their optical properties and the question of their
normality, especially their past/present interaction state. The analysis of the
properties of NGC~7318B are particularly relevant for the understanding
the dynamical state of SQ because they suggest that it is a recent arrival
in the system (see Moles, Sulentic \& M\'arquez 1997; MSM).
The observations are presented in section 2 and analyzed
galaxy-by-galaxy in section 3. The results are combined and summarized
in section 4.

\section{Observations and Data Reduction}

The data were obtained at the Calar Alto (Almer\'{\i}a, Spain) and Roque de los
Muchachos (La Palma, Spain) observatories. CCD images were obtained at
the prime focus of the 3.5m telescope (BVR bands, with a TEK 1024$\times$680
CCD) and with the EOCA 1.5m telescope (BR bands, with a Thompson
1024$\times$1024 CCD) at Calar Alto. The 3.5m telescope was also used with
the Twin Spectrograph (same detector as for the images) to obtain long slit
spectra of NGC~7317, NGC~7319, and through the knots seen North of the central
pair (see figure 1). Long slit spectra of the other galaxies were obtained
with the Intermediate Dispersion Spectrograph attached to the Cassegrain
focus of the 2.5m telescope in La Palma, equipped with the IPCS. All the
spectra were obtained using a slit of 1.5 arcsec width. The log of
the observations is presented in table 1.

\begin{table*}
\begin{center}
\caption[]{Log of the observations}
\begin{footnotesize}
\begin{tabular}{l|  c c c c  c c c c c l c }
\\
\hline
\\
Object & Telescope   & Spectral band   & Exp. (s)  & PA &
FWHM(") & \AA/pixel & Date \\
\\
\hline
\\
SQ & 3.5m & Johnson B & 1200 & - & 0.9 & - & 08/20/88\\
SQ & 3.5m & Johnson V & 1000 & - & 0.9 & - & 08/20/88\\
SQ & 3.5m & Johnson R &  700 & - & 0.9 & - & 08/20/88\\
SQ & 1.5m & Johnson B & 5400 & - & 1.2 & - & 10/12/92\\
SQ & 1.5m & Gunn r    & 2100 & - & 1.2 & - & 10/12/92\\
NGC 7317  & 3.5m & 4200\AA - 5500\AA & 2500 & 109 & 1.1 & 0.9 & 10/31/89\\
NGC 7318A  & 2.5m & 3200\AA - 7000\AA & 3000 & 109 & 1.2 & 2.04 & 08/04/89\\
NGC 7318B  & 2.5m & 3200\AA - 7000\AA & 3600 &  27 & 1.3 & 2.04 & 08/04/89\\
NGC 7319 & 2.5m & 3200\AA - 7000\AA & 1900 &  142 & 1.2 & 2.04 & 08/04/89\\
NGC 7319  & 3.5m & 6000\AA - 7200\AA & 2000 & 142 & 1.1 & 1.2 & 08/25/89\\
NGC 7320 & 2.5m &  3200\AA - 7000\AA & 5000 & 134 & 1.5 & 2.04 & 08/02/89\\
\\
\hline
\end{tabular}
\end{footnotesize}
\protect\label{tbl-1}
\end{center}
\end{table*}


The image were de-biased, flat-fielded and calibrated using the appropriate
routines in the FIGARO package. Extinction and calibration stars were
observed in order to calibrate the 3.5m telescope data. Errors in the
standard stars are always smaller than 3\%. We plotted the standard stars
and stars in the frames in a color-color diagram in order to determine the
accuracy of the calibration. The rms of the distance between the frame
stars and the standards was taken as the calibration error.
This turned out to be similar or less than the errors in the standard stars
themselves. Total photometric parameters were determined for each galaxy.
Indeed, the distortions and overlapping of the different galaxies make
those determination rather uncertain and will be discussed case by case.
Total magnitudes were obtained from the curve of growth and total color
indices were derived.  Color maps were produced to look for the presence of
peculiar features.  Azimuthally averaged light profiles were determined
for the galaxies (see \cite{mam96}, for details about the data reduction
and analysis). As we will explain later, we used an elliptical model fitted
to NGC~7318A to deduce its photometric parameters. Once the model was
subtracted, the remaining light distribution was considered to belong to
NGC~7318B.

The spectroscopic data obtained at Calar Alto has a high wavelength calibration
accuracy and this data was used to derive kinematic properties. The accuracy
was estimated from the position of the main sky lines in the frame. The rms
values are always smaller than 0.1~\AA. The La Palma data were also flux
calibrated. Several standard stars were observed for this purpose. The flux
errors, estimated from the comparison of the standards themselves amount to
5\%. The wavelength calibration accuracy is smaller for those data, and
only a limited use of them will be made for kinematic analysis.

We used cross-correlation techniques (\cite{TD79}) to measure the velocity
distribution of the ionized gas, and to determine the central velocity
dispersion of the elliptical galaxies and of the bulges of the
spiral
members of the Quintet. The Mg2 strength was measured using the line and
continuum spectral bands defined in \cite{bfgk84}.

\section{The properties of the individual galaxies in Stephan's Quintet}

The data for the different galaxies are presented separately. We begin with
the brightest galaxy in the group, the low redshift member NGC~7320 and
end with the pair  NGC~7318A,B. An image of the SQ area is presented in figure 1 with galaxy
identification and slit
orientations indicated for spectra presented here. Positions of
identified emission regions are also marked.

\subsection {NGC~7320}

SQ would not have satisfied the compact group selection criteria adopted by
Hickson (1982) if NGC~7320 were not superimposed. In this sense, it is a
compact group by accident. Without the superposed late type Sd galaxy,
the remaining four members would form a quintet with fainter NGC~7320C (it
does not affect the isolation of the catalogued SQ=HCG92 because it is more
than three magnitudes fainter than NGC~7320). Other  field galaxies are
sufficiently bright and nearby that this wider quintet would not be
sufficiently isolated to pass into the (Hickson 1982) compact group catalog.
So the galaxies physically associated in this system form a
somewhat less compact/isolated system than most Hickson groups.
The core quartet is, of course, highly concentrated with a mean galaxy
separation of 35 kpc. It is likely that many more groups of this kind
escaped inclusion in the HCG.

We measure a redshift cz = 801 ($\pm$20)\kms for NGC~7320 that is
consistent with previous values. The corrected blue magnitude, taking into
account galactic and internal extinction (A$_i$ = 0.40 and A$_g$ = 0.35,
respectively; RC3) is  B$_T$ = 12.35 (12.53 in Hickson et al. 1989).
The galaxy is rather blue with (corrected) indices (U$-$B)$_T$ =
$-$0.11 (RC3), and (B $-$ V)$_T$ = 0.46, consistent with its late
morphological type. Schombert et al. (1990) report a similar B$-$V color
in their grid photometric study where N~7320 is also misidentified as one
of the accordant redshift components in SQ.


No emission lines other than marginal [OII] are detected from the weak central
bulge (see figure 3). The long slit data show no or very weak emission except in
localized regions. Its distribution along the major axis of the galaxy
is asymmetric (see figure 2a). Emission regions were only detected in the NW
half of the disk. The H$\alpha$ image shown in MSM indicates that this is, in part
an accident of slit positioning. A more symmetric HII region distribution is found with
the brightest condensations, admittedly,  on the NW side. Two of these emission
regions, at 19" and 35" from the center
(AR5 and AR4 in Figure 1) have redshifts (730$\pm$40 and 760$\pm$25
respectively) consistent with membership in NGC~7320. They correspond
to the regions denoted A7 and A8, respectively, in Arp (1973). They show
line ratios (see figure 2b) typical for HII regions (with a metallicity of Z
$\sim$0.5$Z_\odot$, as frequently found for late-type spiral galaxies). The
other three detected HII regions at 54",  60", and 75" from the center
(AR3, AR2 and AR1 in Figure 1) belong to the accordant quartet.


Our photometric data shows no marked internal peculiarities. As discussed in MSM, there
is no clear evidence to suggest that the tail that seems to emerge
from NGC~7320 actually belongs to it. The low redshift H$\alpha$ image presented in that
paper shows no evidence for any condensations in the tail though they are clearly seen
on broad-band images. It is therefore most easily interpreted as a background feature
related to interactions in the accordant redshift quartet. The fact that this tail is
parallel to a second brighter one located towards the NE supports that assumption since
the second tail is clearly not related to NGC~7320.

The relatively undisturbed morphology of NGC~7320 suggest that it would be reasonable
to apply criteria for estimating redshift independent distances. Kent (1981) applied
a $\Delta$V(21cm) vs. log L relation to NGC~7320. It was found to fall ``somewhat''
(meaning far) off the calibration relation in the sense that $\Delta$V is too large for
the derived optical luminosity. Shostak et al. (1984) applied a rotation amplitude vs.
size relation and also found that $\Delta$V is too large for the observed optical diameter.
Bottinelli et al. (1985) report a corrected width of 160 \kms~ while Sulentic and Arp (1983)
give a  value $\Delta$V$\approx$ 200\kms. The latter value agrees with the Westerbork value
of Shostak et al. Thus the latest determination favor a somewhat too large value for the
21cm line width. This may support the contention of Sulentic \& Arp (1983) that the
HI emission in NGC~7320 is peculiar and extended. Low redshift HI synthesis
observations with sensitivity to extended structure could prove interesting.
We have re-applied the Tully-Fisher relation using the calibration given
by Pierce \& Tully (1992). Good agreement is found between B$_T$ and
$\Delta$V(21cm) consistent with a distance r$\le$12 Mpc for NGC~7320.

\subsection{NGC~7317}

The measured parameters for NGC~7317 are presented in Table 2. The galaxy
appears to be a normal E2 system. It was also classified as an
elliptical by \cite{hka89}, but Hickson (1994) assigns an Sa type. The
observed light distribution is affected by a nearby bright star plus (possibly)
a background galaxy which can give the impression of an arc-like structure
resembling a
spiral arm. We derived photometric properties for this galaxy by fitting
ellipses to the inner parts where the starlight has no influence and then
extrapolating the model through the outer regions. The adequacy of the
fitting was judged by examining the residuals of the model subtraction.
Only small amplitude residuals at the noise level were observed, indicating
that the fit was satisfactory. The results are presented in figure 4, where
we plot the intensity, ellipticity and position angle profiles. They appear
typical for a normal elliptical galaxy. The data are given in table 2.
The ellipticity profile shows some structure but this is not unusual especially
for ellipticals belonging to groups (Bettoni \& Fasano 1993).
Finally the  color map presented in figure 5a shows a smooth distribution
of light for NGC~7317 without structure or well delineated components.
Thus NGC~7317 shows no evidence for past or present merger activity.

\begin{table*}[]
\begin{center}
\caption[]{Observed photometric and spectroscopic parameters.}
\begin{footnotesize}
\begin{tabular}{l|  c c c c  c c c c c l c }
\\
\hline
\\
Object & B$_T$ & (B$-$V)$_T$ & (B$-$R)$_T$ & PA & i & cz & $\sigma_c$ & W &
Mg$_2$\\
\\
\hline
\\
NGC 7317   & 14.37 & 1.10 & 1.72 & 115 & -  & 6563 & 215 & -   & 0.28\\
NGC 7318A  & 14.33 & 0.94 & 1.65 & 105 &    & 6620 & 230 & -   & 0.27\\
NGC 7318B  & 13.92 &      &      &  22 & 51 & 5765 & 220 & 145 & 0.22\\
NGC 7319   & 14.23 & 1.00 & 1.68 & 140 &    & 6650$^a$ & -   & 200 & -\\
NGC 7320   & 13.11 & 0.64 & 1.12 & 133 & 59 &  801 & -   & $\leq$100 & -\\
\\
\hline
\end{tabular}
\end{footnotesize}
\protect\label{tbl-2}
\begin{scriptsize}
\noindent

(a) This value corresponds to the systemic redshift, see text.
\end{scriptsize}
\end{center}
\end{table*}
\noindent\null\par


From our spectrum of the central region we measure a redshift cz = 6563~($\pm$29)~\kms in
good agreement with the value reported by \cite{hmhp92}. No rotation is seen along the
slit at our position angle. The central velocity dispersion was measured to be $\sigma$ =
215 ($\pm10$) \kms~ compared to 299 \kms~ reported by Kent (1981). The Mg2 strength is
0.28 which is typical for early-type galaxies. From the photometric and spectroscopic data
we conclude that NGC~7317 is a reasonably normal elliptical galaxy. Application of the
\cite{fj76} relation as revised by \cite{dvo82} gives D $\sim$~65 Mpc, in agreement with
the photometric estimate of \cite{dvo84}.

This leaves open the question of whether NGC~7317 is an active participant
in the dynamical evolution of SQ. There is no internal evidence for a past
merger and no obvious evidence for interaction.  The single strong form of
evidence for active participation in SQ comes from the diffuse halo which
clearly extends from the region of NGC~7318A,B towards NGC~7317. This halo
can be seen on most deep images published over the past 25
years (Arp 1973, Arp and Kormendy 1972, Arp and Lorre 1976 and Schombert et
al. 1990). Sandage and Katem (1976) suggested that much of the faint emission
could be due to high latitude galactic cirrus. While this
interpretation is probably true for much of the structure seen far
from SQ, there is unambiguous diffuse structure associated with
the group that represents evidence for dynamical evolution. We
identify two components to the SQ halo: 1) a redder component
that is particularly strong to the NW of NGC~7317 and 2) a bluer component
associated with the tidal tails extending towards the SE. The former
component clearly links NGC~7317 with the rest of the accordant galaxies.

We estimated the extent and integrated luminosity of the older halo
component using IDL tasks after standard reductions and flat fielding.
We removed the foreground stars which are numerous in the SQ region. We
determined the boundary of the halo by smoothing the R band image with
various mean and median filters. Figure 5 shows the adopted boundary
(heavy solid line) of the diffuse light chosen as the last contour level
above the noise (approximate surface brightness level 25.6 mag/('')$^2$).
Note that the
halo boundary was interpolated across NGC~7320 and the blue tidal tails.
The halo becomes very nonuniform north of NGC~7318AB where active star
formation related to new activity is observed. In order to make a
conservative estimate of the halo luminosity we chose the region between
NGC~7317 and 7318A as ``typical'' with mean surface brightness 24.4 mag
arcsec$^{-2}$. This average level was multiplied by the total area inside
of the adopted boundary . Using our assumption of
constant surface brightness we obtain M$_{halo}$= -20.9 $\approx$ L$_\ast$
$\approx$ the average luminosity of an SQ component.


The extension and surface brightness of this
diffuse light rules out the possibility that is produced by overlap of
individual galaxy haloes, especially NW of NGC~7317. We note that NGC~7317 is
not centered in the diffuse light that extends in its direction. This may
mean that it has a considerable transverse component in motion relative to
the rest of the group. This is almost required for the group to avoid
collapse because the relative radial velocities of NGC~7317, 18A and 19
are very small.

\subsection {NGC~7319}

This SBc galaxy shows the strongest evidence for tidal disturbance. It shows a
strong bar at PA = 142 which is also the position angle of the major axis.
The arms show no evidence for any emission regions or dust lanes.
The general aspect of the galaxy with its tails and arcs has been described
in the literature (\cite{ako72, sch90}). Our photometric results
are  given in table 2. A blue (B$-$V= 0.57) tidal tail
extends towards the SE from the edge of NGC~7319. The connection to NGC~7319
is not unambiguous as it shows no smooth connection to the spiral arms in
that galaxy. The tail is parallel to and brighter than the one that
passes behind NGC~7320. Both tails curve in the direction of NGC~7320C. At the
same time, HI mapping (Shostak et al. 1984) shows that NGC~7319 has been completely
stripped of its neutral hydrogen. H$\alpha$ images (Arp 1973 and MSM) show almost no
emission from the galaxy as well. Recent CO mapping of NGC~7319 (Yun et al.
1997) show only a few small molecular clouds that appear unrelated to the
spiral structure. They match very well dust patches seen on deep optical
photographs and may well represent stripped material above the disk.  All those
evidences support the view (Shostak et al. 1984) that NGC 7319 has suffered a strong
collision/encounter that has efficiently stripped its ISM. That tail points towards
NGC~7320C as the intruder responsible for the stripping event (Shostak et al. 1984; MSM)

Despite the evidence for strong dynamical evolution the galaxy retains reasonably
symmetric spiral structure. It satisfies the Tully-Fisher relation
when placed at its redshift distance.

NGC~7319 is a well known Seyfert 2 galaxy (see \cite{du87}) with emission lines
that show strong blue wings. The spectrum obtained at Calar Alto shows
extended emission especially towards the NW side. The bulk of the emission
however is not H$\alpha$ but [NII]. Figure 7a presents the observed
$H\alpha/[NII]\lambda6583$ ratio along the slit. It can be seen
that at some points only the [NII]$\lambda$6583 line was  detected and, in general, the
line ratio is smaller than 1. This suggests that residual gas in
NGC~7319 is shocked --another manifestation of the past collision.

The kinematic data are presented in figure 7b. The velocity distribution is
irregular (PA= 142$^\circ$). The observed velocity gradient,
amounting to $\sim$ 200 \kms, suggests that this may also be the kinematic
major axis. The value we obtain for the redshift at the
position of the active nucleus is 6770 ($\pm25$)~\kms~ in good agreement
with Hickson et al. (1992). However the kinematic center deduced from
symmetrization of the central part of the observed velocity distribution, is
slightly offset
($\sim$4") with respect to the active nucleus. The latter coincides with
the continuum maximum while the kinematic center rests on the
fainter knot shown in figure 6b. Therefore the systemic redshift would be
V= 6650 \kms rather than the value measured for the active nucleus.



An interesting result comes from the color map shown in figure 6. A well
delineated structure appears in the very central region, in a direction
perpendicular to the bar, where the brightest (and reddest) knot coincides
in position with the active nucleus. That structure is coincident with the
jet-like continuum radio source reported by van der Hulst and Rots (1981)
and with the extended emission-line region. The large scale
outflow recently reported by Aoki et al. (1996) also extends along
the same direction (see figure 6b). A larger and broader red feature
extends from the NGC~7319 nucleus and towards the nucleus of NGC~7318B. This
feature was probably detected on the high redshift
H$\alpha$ image shown in MSM. Arp (1973) had earlier reported
detection of
this structure in H$\alpha$. Perhaps we are seeing the last remnants of the
shocked emission associated with the passage of NGC~7320C through SQ--that
resulted in the complete stripping of HI from NGC~7319.

\subsection {NGC~7318A}

The overlapping light distributions of  NGC~7318AB make it difficult to
assign a morphological type to each galaxy. Regarding component A, it is
not even a simple matter to decide whether it is an elliptical or a late
type spiral. This last possibility is motivated by the presence of arm-like
structures South and East of the central body that could be associated with
it (see \cite{ssa84}). The type given in RC3 is .E.2.P., but it was
classified as Sc by \cite{hka89}, and as SB0 by Hickson (1994).

In order to help clarify the morphology of NGC~7318A we have followed the
RC3 by considering the galaxy to be an elliptical, and tried to
fit an elliptical light distribution to it. No objectively defined
criteria were used, except that the residuals after model subtraction should
reasonably attributed to the companion galaxy NGC~7318B.
The main parameters, which are presented in table 2, were measured from the
fitted model. Comparison with values for NGC~7317 indicate that they are
rather similar galaxies.  The result of the model fit and subtraction is
shown in figure 8. We notice that there is no sign of a bar and little
evidence, if any, for a disk but there is complex residual structure. While
not unprecedented in an elliptical galaxy, it could be evidence for a past
interaction and/or merger event, related to the (hypothetical) earlier passage
of NGC~7320C.

We measured a redshift of 6620 ($\pm$20) \kms, in excellent agreement with
Hickson et al. (1992) and slightly lower than  RC3. We could not detect
any rotation of the stellar component at the observed PA. On the other hand,
the spectroscopic results strongly reinforce the view that NGC~7318A is an
early type galaxy. The central spectrum, shown in figure 3, is much redder
than that for NGC~7318B. The Mg2 index is 0.27 mag which is very similar
to the value for NGC~7317. The  central velocity dispersion is
$\sigma$ = 230 ($\pm$ 10) \kms~ again lower than the value reported in
Kent (1981). Application of the Faber--Jackson relation yields a most probable distance
of $\sim$65 Mpc, consistent with its redshift and the value obtained
by \cite{dvo84}.

Bushouse (1987) reported detection of modest H$\alpha$ emission from the
nuclear region of NGC~7318A. However, higher resolution photometry by Keel
et al. (1985) reports only a more sensitive  upper limits to that emission.
Our long slit data show no emission. In fact, only two emission
regions were detected along the slit, located 25" and 34" East of the
nucleus (BR1 and BR2 in figure 1). They correspond to regions B7 and B8
in Arp (1973). The first belongs to NGC~7318B. Only [OII]  was detected in
the second region  at a redshift  $\sim$6300 \kms~ suggesting that it
could be part of the ISM stripped from NGC~7319. The same comment can be made
for emission region AR3 with V= 6460$\pm$40 \kms.


\subsection {NGC~7318B}

Subtraction of the elliptical  model for NGC~7318A, leaves a reasonably well
defined barred spiral galaxy (figure 8). NGC~7318B is classified  SBbc in
RC3. Hickson et al. (1989) assigned it an Sbc classification. The photometric
parameters are given in table 2. No emission was detected in the  long slit
spectrum at PA = 27 (i.e. along  the bar). This is compatible with the idea
that it is an old population II bulge+bar. The central velocity dispersion
is 220 ($\pm$10)~\kms, which is not unusual for the bulge of an early-type
galaxy. The Mg2 strength amounts to 0.22  which is smaller than the
values found for NGC~7317 and NGC~7318A.

Our photometric results support the idea that NGC~7318B is still relatively
unperturbed as it still shows a well defined spiral pattern. Its high relative
velocity ($\Delta$V$\sim$10$^3$ \kms) and the radio/X-ray evidence for an
ongoing collision on the eastern edge of this
galaxy (Pietsch et al. 1997; van der Hulst and Rots 1981) strongly support the view
(Shostak et al 1984) that NGC~7318B is now entering the group and for the first time. This
collision is shocking the IGM of the group along with the ISM of the
collider. The interface can be well seen on the high redshift H$\alpha$ image
shown in MSM. The idea that NGC~7318B is
still a reasonably intact disk galaxy is supported by our reinterpretation of
the HI data reported by Shostak et al. (1984). We suggest that the two HI
clouds detected at $\sim$5700 \kms~ and $\sim$600 \kms~  actually belong to
NGC~7318B. When the HI profiles of the two clouds are superposed, the result
is a double horn profile. The 2D HI distribution shows a hole in the center
that is coincident with the central bulge of the galaxy as is frequently
found in spiral galaxies (figure 9). The central redshift of the combined
profile agrees with the
value of 5765 ($\pm$~30)\kms~ derived from the  optical spectrum.
The (uncorrected) amplitude of the implied rotation amounts to 145 \kms.


Our successful model subtraction of NGC~7318A suggests that all of
the spiral structure in the vicinity of 7318AB can be ascribed to component B.
The outer arms are somewhat chaotic and the region north of 7318B is confused
at the intersection of the shock front and the northern spiral arm. Some
of the detected HII regions are almost certain members of NGC~7318B, in particular
the two emission regions detected on the N edge of NGC~7320 (AR1
and AR2 in Figure 1) with V= 5540$\pm$40 and 5800$\pm$70 \kms and
BR1 located 25" E from the center of NGC~7318A with V = 5850 \kms.
Finally two of the detected emission knots North of the central pair
(marked K1 and K2) could also be external HII regions  associated with
NGC~7318B. Both have the same redshift cz$\sim$ 6020 \kms, and correspond
to regions B13 and B11 in Arp (1973). These velocities are consistent with
our interpretation of the HI in the sense that the higher velocities are
towards the north side of the galaxy.

If we accept that the atomic gas component of the NGC~7318B is still
largely in place we can apply the Tully-Fisher relation to it.
The value we obtain from the (corrected) observed parameters is $\sim$60 Mpc,
similar to the values we obtained for the other high redshift members of
the group.

\section{Final remarks}

The observational evidence suggests that the SQ galaxies preserve a well defined
identity in spite of strong signs of present and past interaction. NGC~7317 and 18A
show properties of elliptical galaxies. The data presented here has been used to
estimate their distances through the use of the Faber \& Jackson relation.
The optical and HI data on NGC~7318B confirm that it is a relatively undisturbed
spiral. The Tully-Fisher relation was applied to find its distance. The
other two spirals in the group also satisfy the Tully-Fisher relation at their respective
redshift distances.

The signs of interaction are mainly the two tidal tails on the SE side of SQ,
both pointing to NGC~7320C, the diffuse light surrounding
the group, and the structure seen East of NGC~7318B. The data show that NGC~7318B
still has a recognizable spiral structure and an HI disk, indicating that its
interaction with the other three accordant members of the group has begun only recently.
A strong shock front appears to mark the interface of the collision. This is best
seen in radio continuum and X-ray light but it is also visible in the high redshift
H$\alpha$ image shown in MSM. Some of the
detached emission knots were detected in our slit spectra. We suggest that knots AR3
(cz$\sim$6460 \kms), BR2 (cz$\sim$ 6460 \kms) and K3 (cz$\sim$ 6680 \kms) represented
some of this new proto-halo material. The first two show very weak H$\alpha$ and well
detected [OII]$\lambda$3727. The region K3 corresponds to region B10 in Arp (1973). It
shows a high excitation
spectrum, with [OIII]$\lambda$5007/H$\beta$ = 2.6 and
[NII]$\lambda$65847/H$\alpha$ = 0.11. These knots belong to a well defined
structure in the intergalactic medium of the Quintet; the shock interface.
The line ratios in knot K3 indicate a rather high excitation and
therefore a young age. Since these ratios are fully consistent with
photoionization by stars we conclude that there is some very recent star
formation in the shocked zone. The gas is being shock heated  and the
emission regions represent material that may already be cooling at optical
wavelengths.

In summary, the existing data suggest that SQ is  composed of: 1) a foreground dwarf
spiral galaxy (NGC~7320), 2) an interacting triplet composed of one spiral (NGC~7319) and
two ellipticals (NGC 7317, 18A), and 3) a recently arrived spiral (NGC~7318B) now entering
the group for the first time. The long structure seen east of the galaxy would witness
this event, whereas the tidal tails support the idea that NGC~7320C probably collided with
the group at least twice in the past. The detected diffuse light detected in the SQ area,
constitutes an argument for the long term dynamical evolution of the group.
In MSM we synthesize these observations and attempt
to relate the implications of SQ to the class of compact groups.

\vskip 2truecm

{\sl Acknowledgments.}

This research was partly supported by DGICYT (Spain) grant PB93-0139. We
acknowledge
the support of the INT and the CAT for the allocation of observing time. The
Isaac
Newton Telescope on the Island of La Palma is operated by the Royal Greenwich
Observatory at the Spanish Observatorio del Roque de los Muchachos of the
Instituto de
Astrof\'{\i}sica de Canarias. We acknowledge J. Masegosa and A. del Olmo for
their participation in the observing runs. I.M. acknowledges financial
support from the Ministerio de Educaci\'on y Ciencia through the grant
EX94 08826734.

\clearpage

{\bf Figure Captions}
\vskip 0.5truecm

Figure 1. R band image of the Quintet area taken with the 1.5m telescope in Calar Alto.
North is at the top and East to the left. The slit positions for which detached emission
knots were detected are shown. Identification is given for those knots
\vskip 0.5truecm

Figure 2. (a) The H$\alpha$ (dashed line) and continuum (solid line) flux distributions in
NGC~7320 along the slit. (b) Spectrum of its brightest detected HII region (AR4)
\vskip 0.5truecm

Figure 3. The spectra of the bulges of NGC~7318A,B, and NGC~7320
\vskip 0.5truecm

Figure 4. Photometric profiles of NGC~7317, (a) the intensity, (b) the ellipticity, and
(c) the position angle
\vskip 0.5truecm

Figure 5. Contour image of Stephan's Quintet. The solid and dashed lines represent the old
and new halo, respectively (see text)
\vskip 0.5truecm

Figure 6. (B-R) color map of the galaxies in the Quintet. Darker is bluer. The two inserts
show (a) NGC~7317, and (b) NGC~7319. The radio contours given in van der Hulst and Rots
(1981) have been plotted on the map of the nuclear region of NGC~7319
\vskip 0.5truecm

Figure 7. (a) The distribution of the H$\alpha$/[NII] line ratio in NGC~7319 along the
slit. (b) The velocity data along the slit. The position of the active nucleus is marked
with AN
\vskip 0.5truecm

Figure 8. B-band image of the pair NGC~7318A,B after subtraction of an elliptical fit to
component A
\vskip 0.5truecm

Figure 9. Reconstruction of the HI profile of NGC~7318B from the data reported by Shostak
et al. (1984)

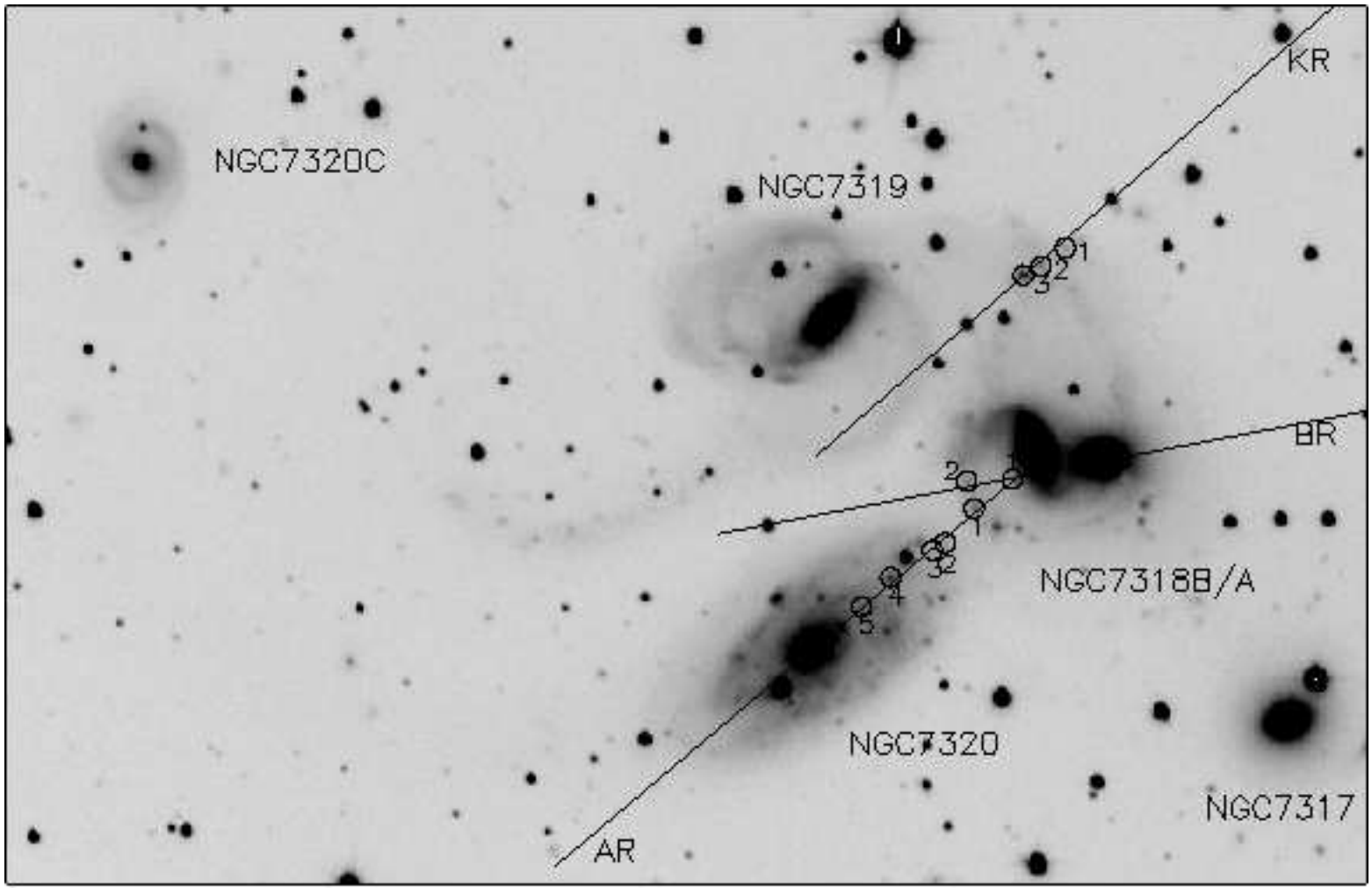
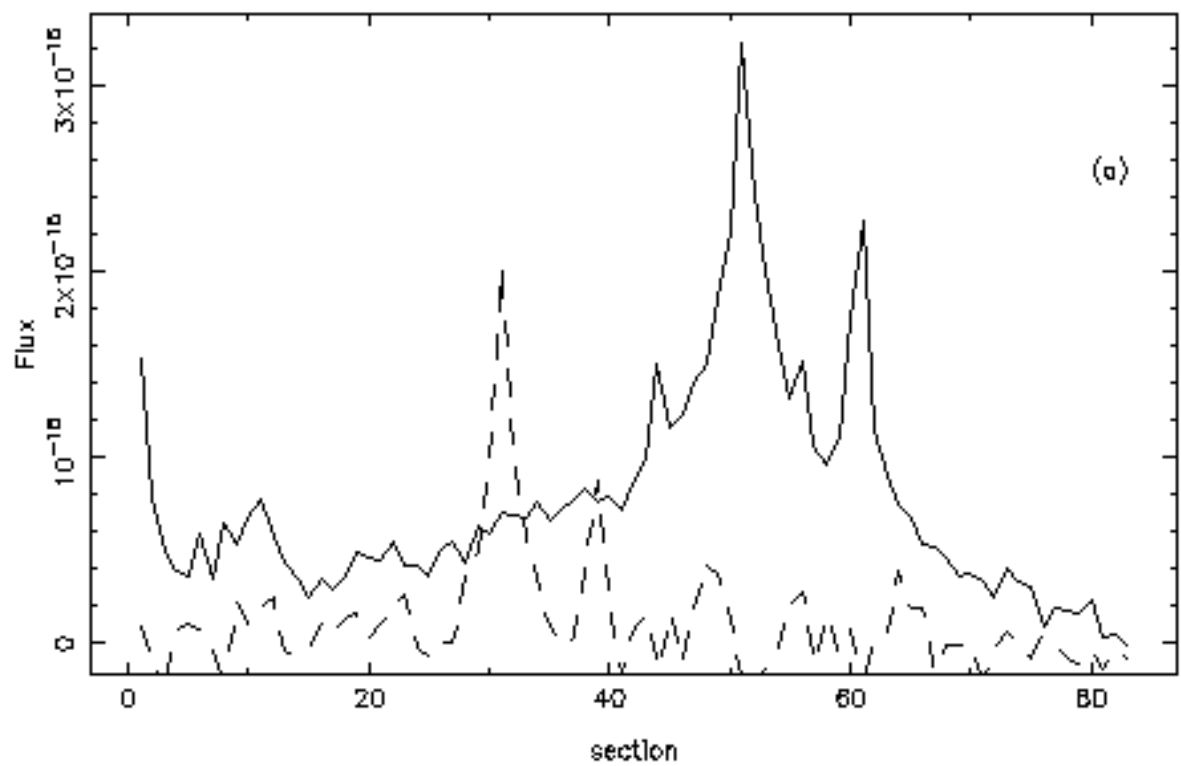
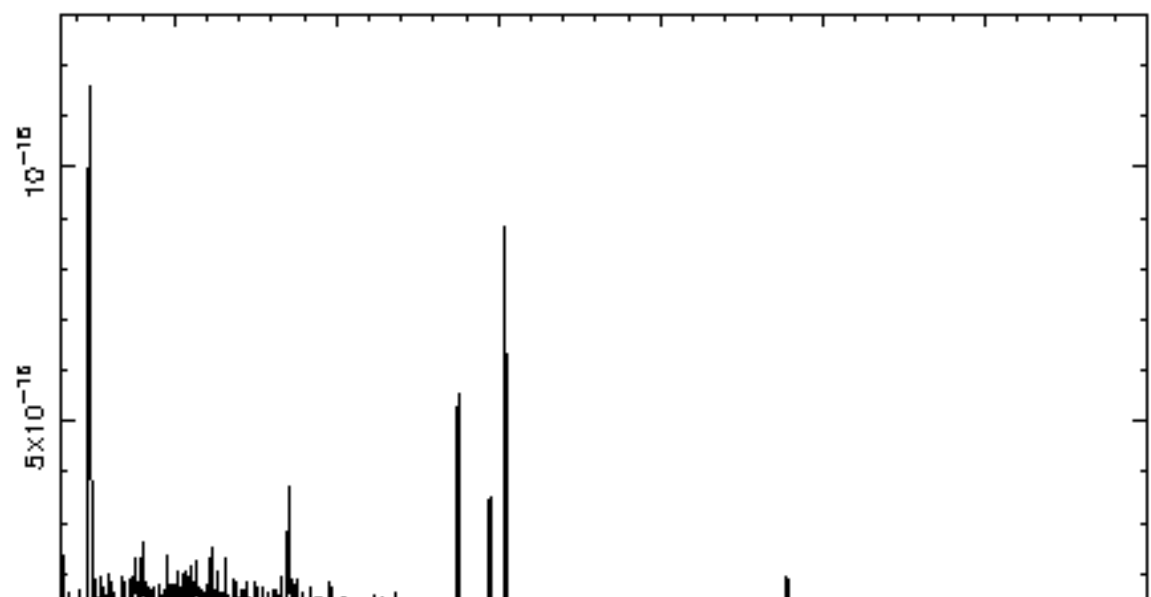
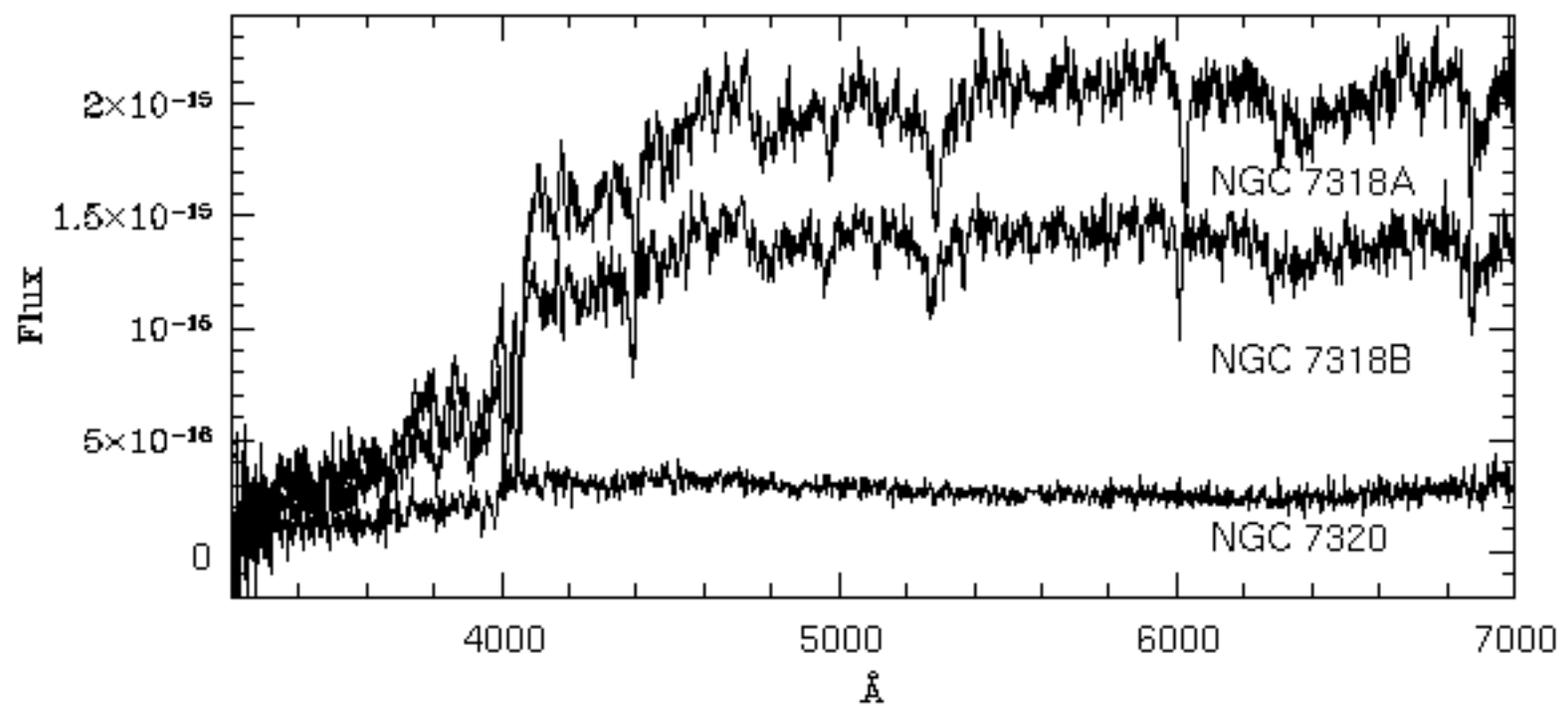
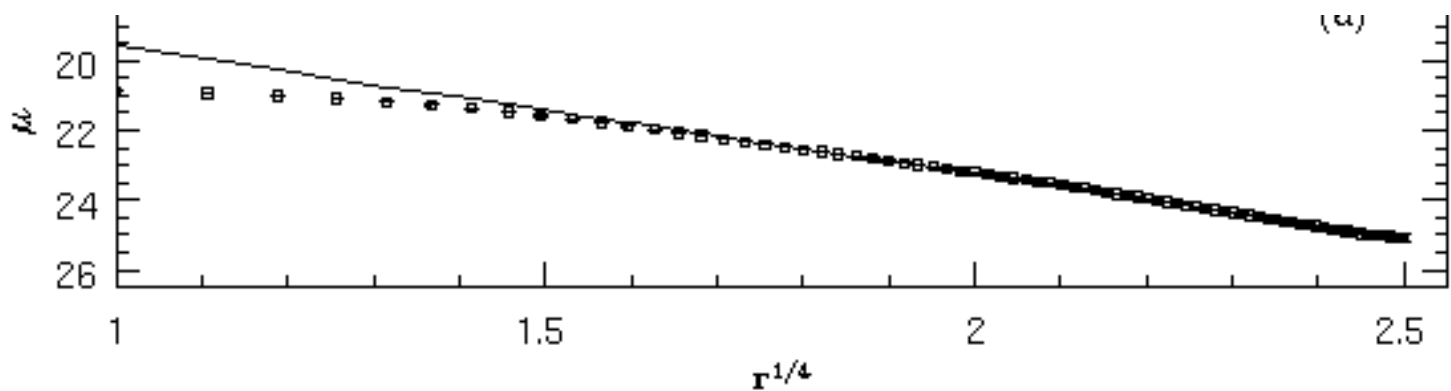
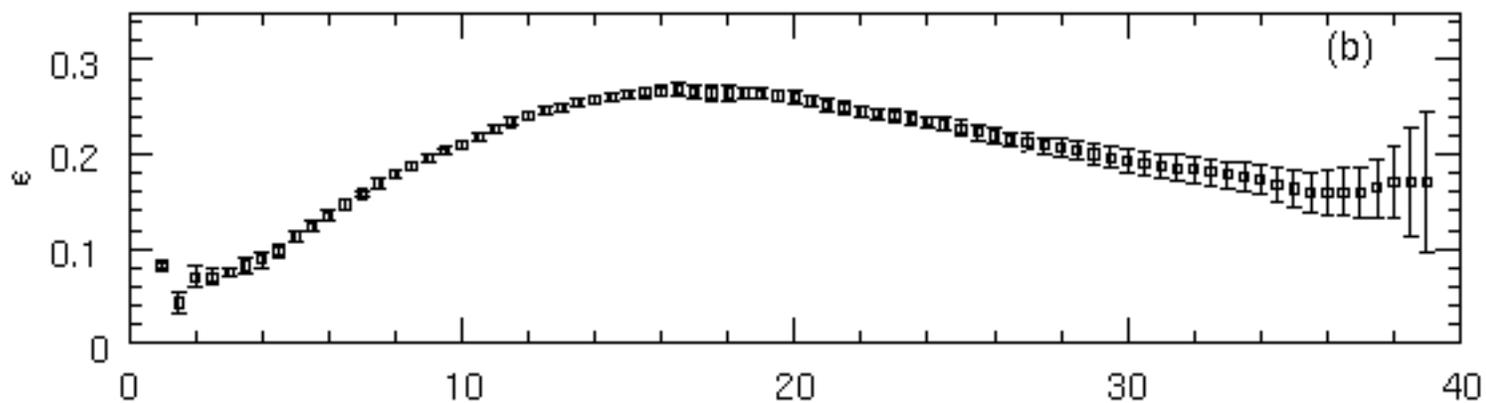
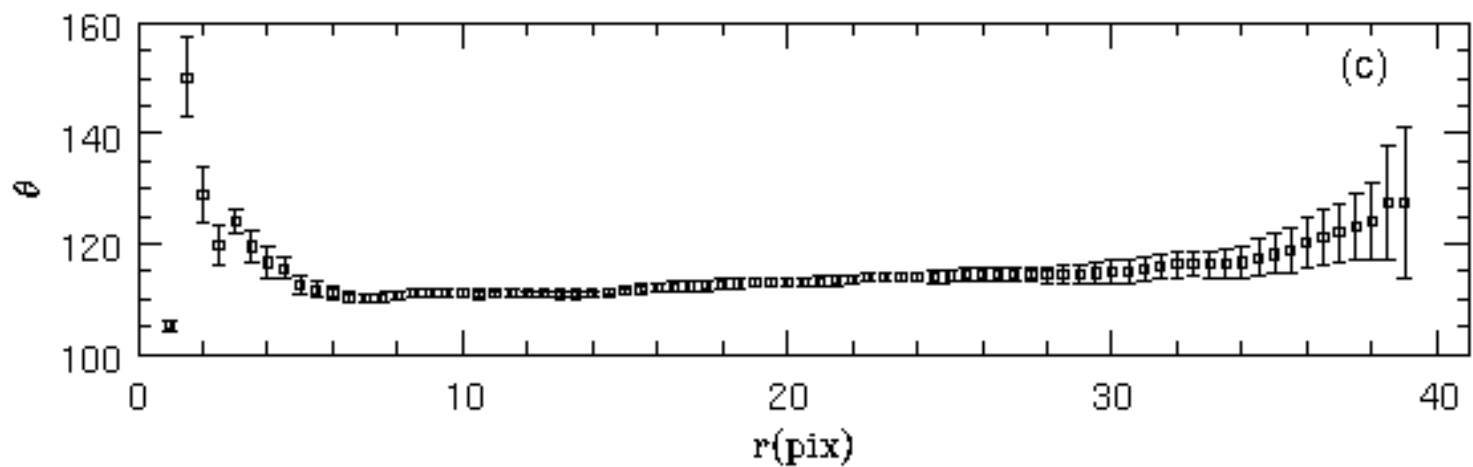
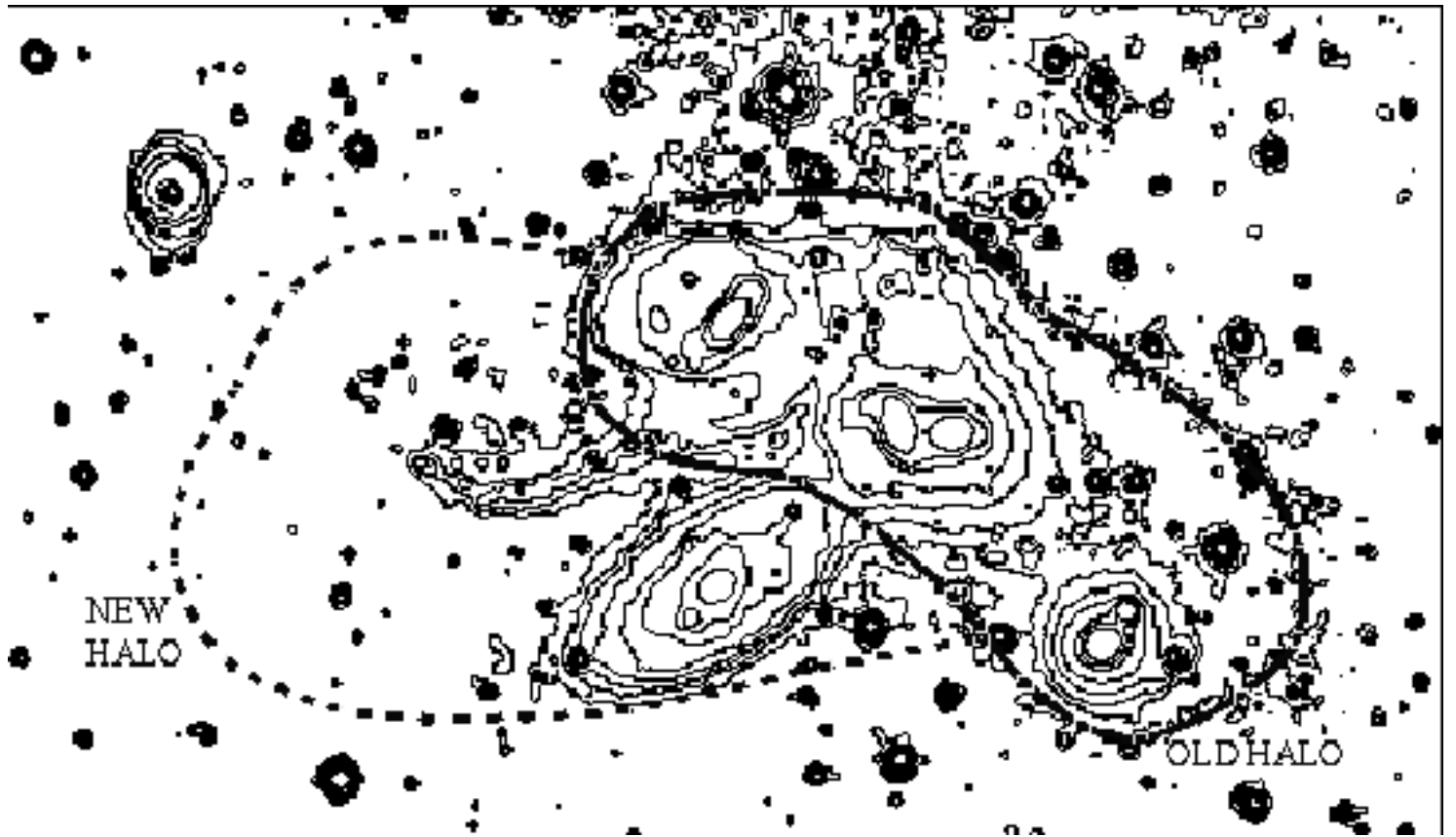
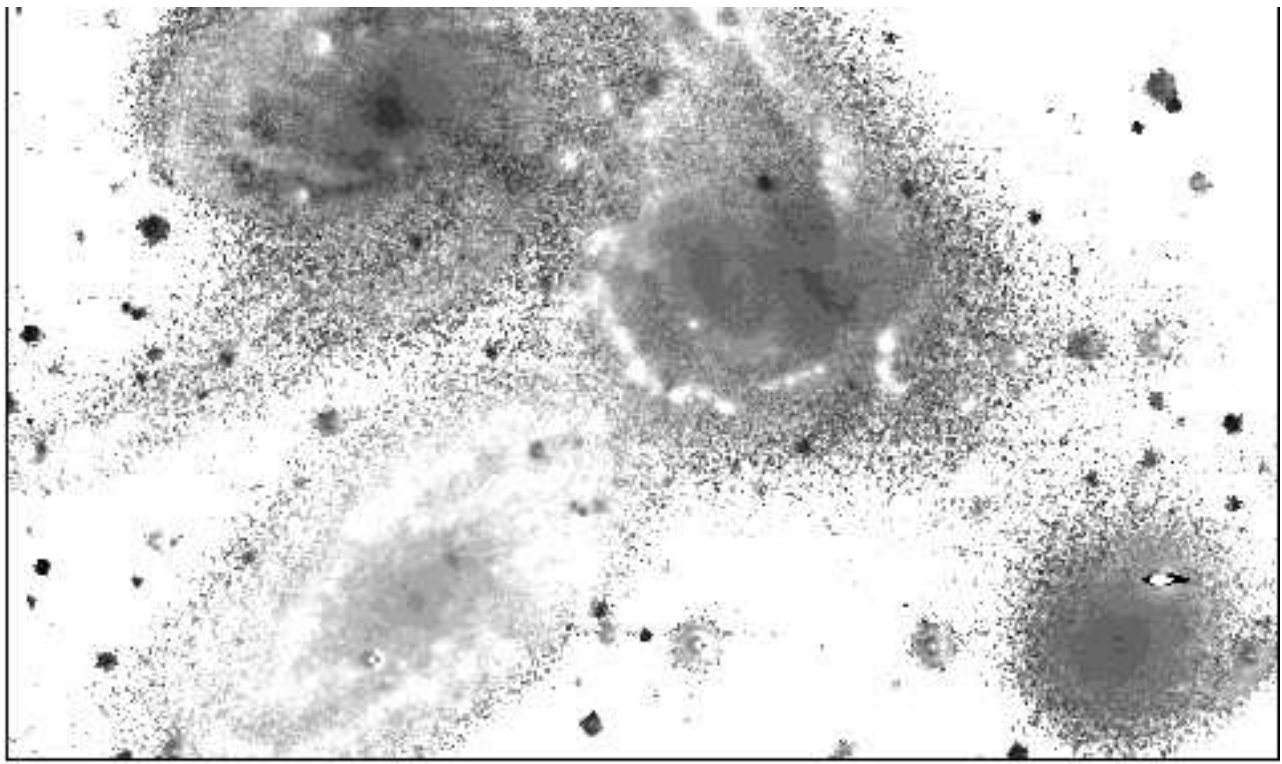
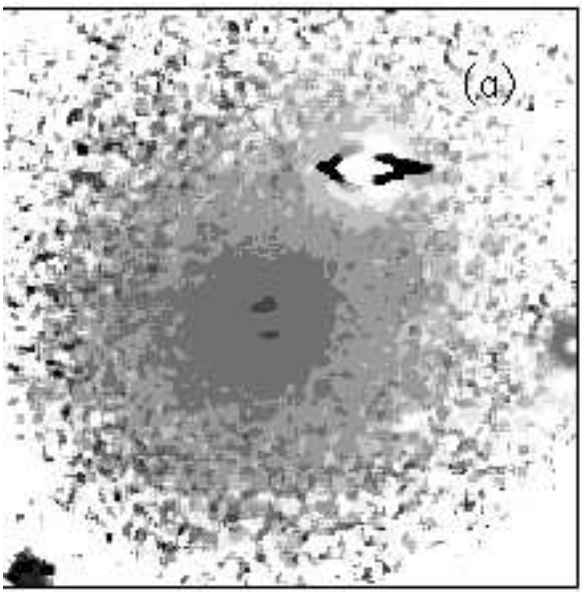
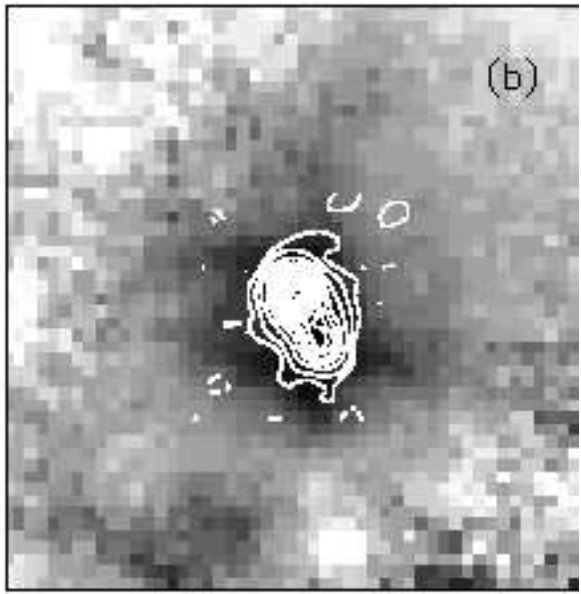
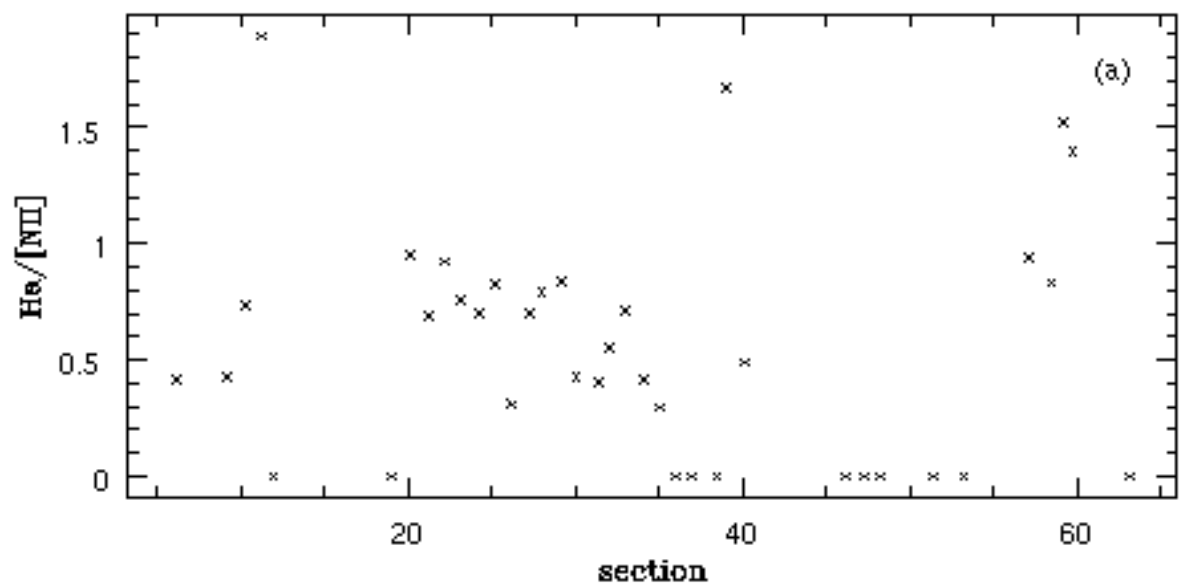
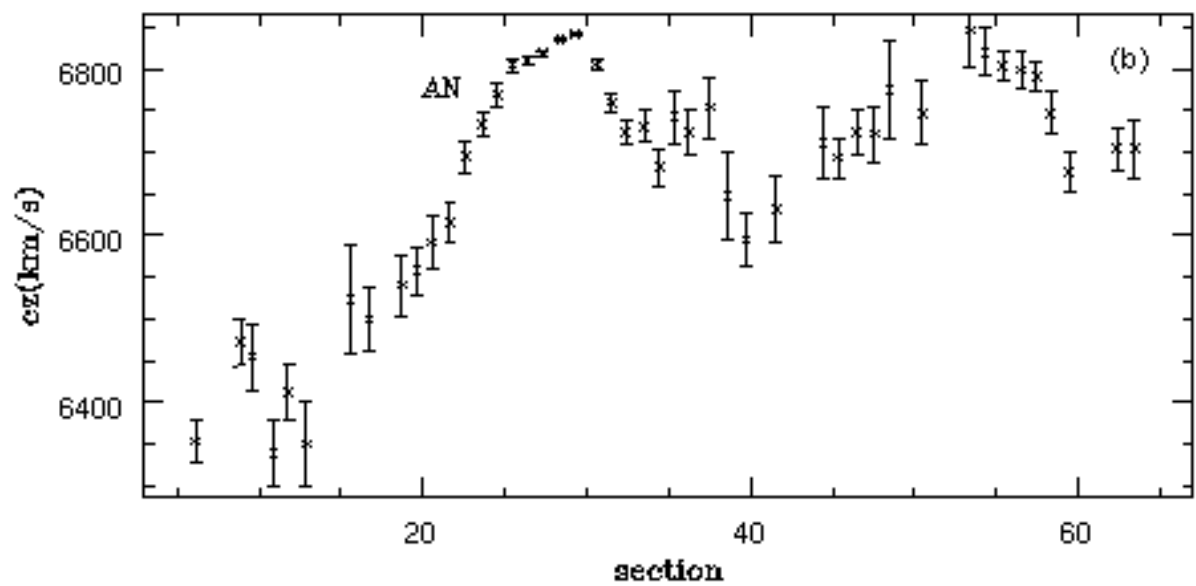
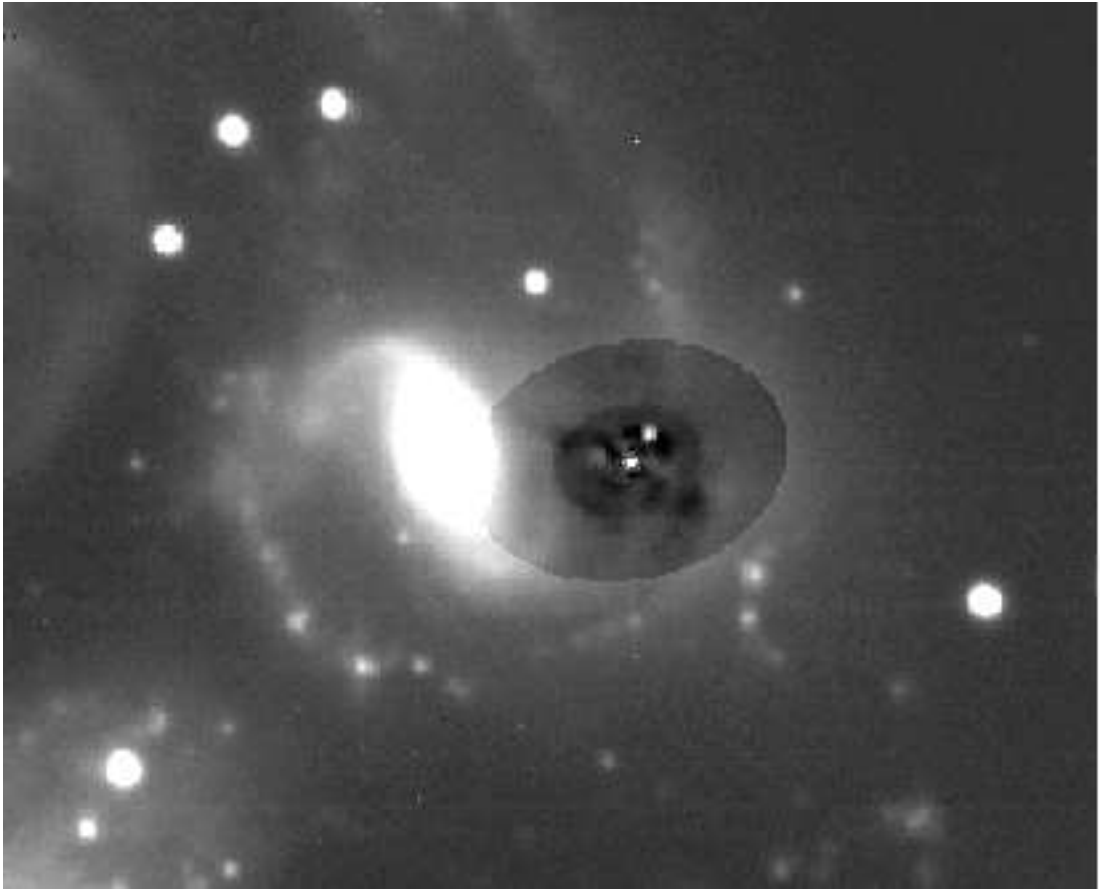
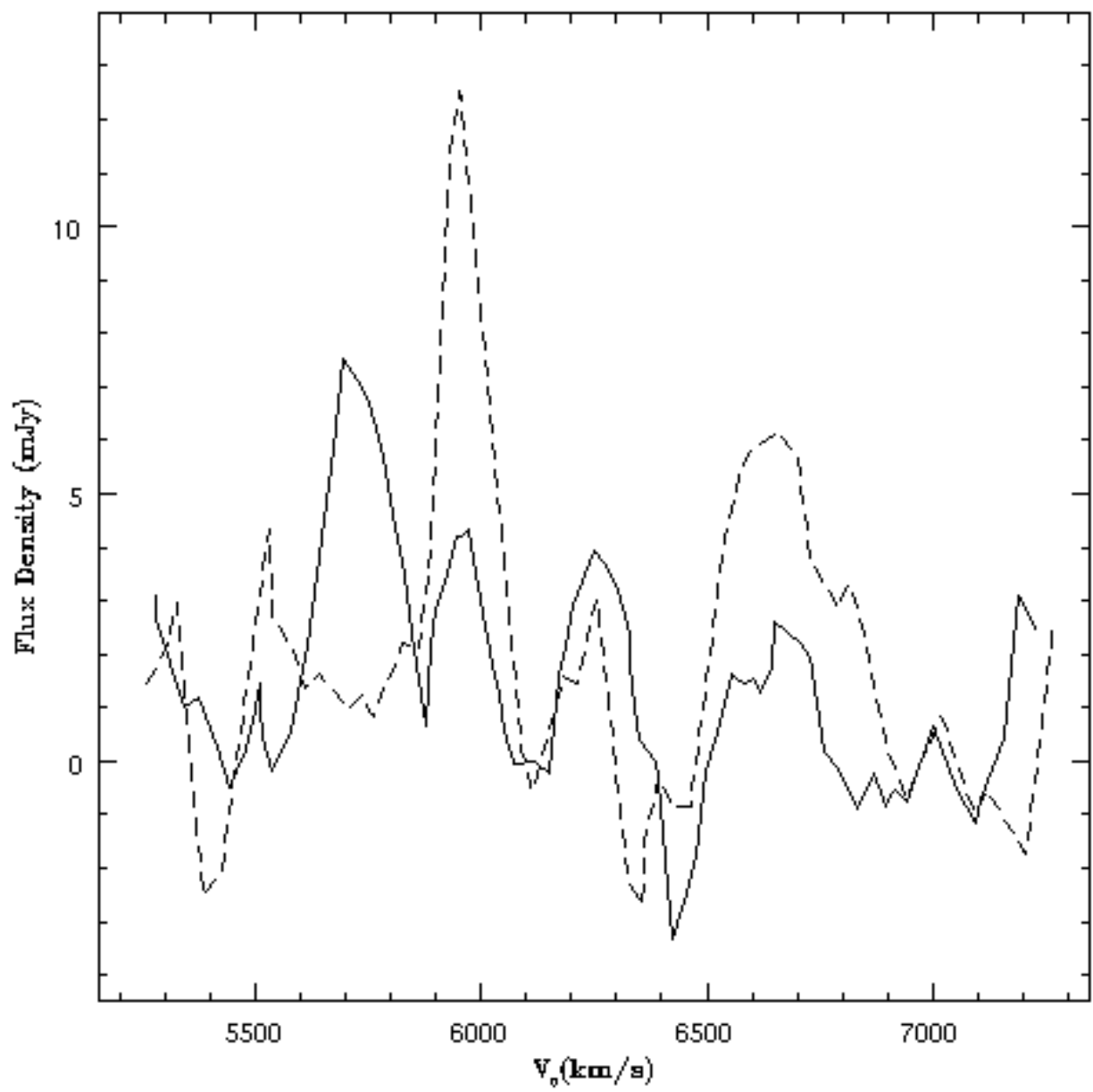
\end{document}